# An Efficient Machine Learning Approach for Accurate Short Term Solar Power Prediction

Mr. Shaktinarayana Mishra
*Department of ECE*
*Gandhi Institute of Technology and Management*
Bhubaneswar,India
shakti.n.mishra@gmail.com

Dr. Lokanath Tripathy
*Department of EE*
*College of Engg. & Technology*
Bhubaneswar,India
lokanath@cet.edu.in

Dr. Prachitara Satapathy
*Department of EE*
*College of Engg. & Technology*
Bhubaneswar,India
prachitara.satapathy@gmail.com

Prof (Dr.) P.K. Dash
*Department of EE*
*Siksha 'O' Anusandhan*
Bhubaneswar,India
pkdash.india@gmail.com

Miss Nitasha Sahani
*Department of Electrical & Computer Enggineering*
*Virginia Tech.,* Blacksburg, USA
nitashas@vt.edu

*Abstract*— Solar based electricity generations have experienced a strong and impactful growth in the recent years. The regulation, scheduling, dispatching and unit commitment of intermittent solar power is dependent on the accuracy of the forecasting methods. In this paper, a robust Expanded Extreme Learning Machine (EELM) is proposed to accurately predict the solar power for different time horizon and weather condition. The proposed EELM technique has no randomness due to the absence of random input layer weights and takes very less time to predict the solar power efficiently. The performance of the proposed EELM is validated through historical data collected from National Renewable Energy Laboratory (NREL) through various performance metrics. The efficacy of the proposed EELM method is evaluated against basic ELM and Functional Link Neural Network (FLNN) for 5 minute and 1 hour ahead time horizon.

## I. Introduction

Solar generations is preferred over other renewable sources as it is sustainable and decarbonized. Further it can be supplied to local loads to minimize the power losses with negligible operation and maintenance costs. Nevertheless, solar power is intermittent and stochastic in nature, It is due to its dependence on unpredictable meteorological and weather conditions like cloud cover, solar radiation and temperature. So, to overcome this drawback, methods of solar power forecasting are adopted for management of electricity grids and energy market efficiency. Short-term forecasting uses localized approach in order to assure grid stability and provide estimation of future solar power generation values.

In recent years, many types of solar power forecasting methods have been implemented [1,2]. Depending on current geographical and meteorological data, mathematical models are used to predict solar power under the numerical weather prediction (NWP) model. The time series methods [3] used cannot detect the high fluctuations in power generation, particularly for cloudy days. Statistical models [4,5] like autoregressive moving average (ARMA), autoregressive integrated moving average (ARIMA) and multiple regression require proper detectors to collect varying weather parameters and are inefficient in short term prediction of non-linear time series accurately. Artificial neural network (ANN) methods [6-8] need complicated structures to detect non-linear function approximation and are quite time consuming owing to training procedures. The fuzzy logic method shows design and computational complexity [9].

The extreme learning machine (ELM) approach is an efficient method to predict the solar power which is faster as compared to other methods. There are various hybrid prediction methods [10-13] which performs better as compared to the single approach. Here, an extended ELM approach is discussed to predict the solar power in which, the basic ELM is combined with a non-linear expansion block [14] called (FEB) to add more non-linearity to the input of the ELM.

The following paper is organized as follows. After a brief introduction in section 1, different proposed techniques for solar forecasting are formulated in section 2 including the proposed model. Section 3 discuses the results including various performance metrics. Lastly, section 4 concludes the paper.

## II. Proposed Techniques

### A. Functional Link Neural Network (FLNN)

A FLNN is an extended version of neural network and uses a functional expansion link to modify inputs [14]. These inputs are further modified by weights then summed and applied to a particular function to calculate the output. During the training, the weights are continuously adjusted and updated to obtain the defined accuracy. The presented FLANN is a single layer neural network in which trigonometric functions are used to expand the inputs with more non-linearity. The structure of the presented TFLANN model is shown in figure 1. Here each individual input '$x_j$' present in the input pattern is expanded using trigonometric functions with order '$p$' as:

$$1, x_j, x_j^2, \sin(\pi x_j), \cos(\pi x_j), \sin(2\pi x_j) \\ \cos(2\pi x_j), ... \sin(p\pi x_j), \cos(p\pi x_j) \quad (1)$$

The solar data patterns are created by applying a sliding window to the original data. Let the model have '$S$' number of data patterns where each data pattern contains '$n$' number of inputs and '$m$' of output. Let '$S_{trn}$' is the total number of training pattern and '$S_{tst}$' is the total number of testing pattern. For each individual input ($x_j$) of '$i^{th}$' input pattern ($X_i$) the expanded inputs are given in equation (2) where '$ei$' is the number of inputs after expansion.

$$z_1^i = 1, z_2^i = x_j, z_3^i = x_j^2, z_4^i = \sin(\pi \times x_j),$$
$$..... z_{ei}^i = \cos(\pi \times x_j) \qquad (2)$$

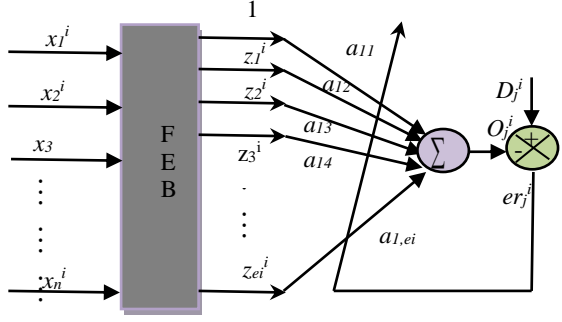

Figure 1. Architecture of Trigonometric FLNN

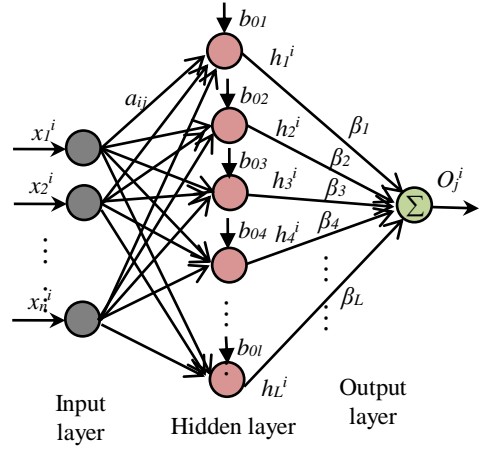

Figure 2. Structure Extreme Learning Machine

After the expansion the predicted value (*O*) for '$i^{th}$' input pattern of is calculated by using the equation (3).

$$O_i = \sum_{k=1}^{ei} \left( a_{1,k} \times Z_k^i \right) \text{ where, } k = 1,2...ei \qquad (3)$$

For the presented study, to predict the six number data past five number of solar power data are considered in each pattern and learning rate for FLNN is kept as 0.01.

The main flaws of this model are need of the proper size historical dataset, proper choice of varying network parameters like learning rate and repeated weight adjustments owing to more computational time. These drawbacks can be taken care of with the efficient ELM algorithm which is much simpler and faster due to the absence of weight adjustment. The increased non-linearity by adding biases and other non-linear activation functions in ELM can handle non-linear data for efficient solar power prediction in less computational time.

*B. Extreme Learning Machines (ELMs)*

The Extreme learning machines are single hidden layer feed forward networks with very fast learning mechanism and thus can be applied for solar prediction [15,16]. In ELMs, the output weights are calculated analytically by the least square solution which makes it simple, fast and more accurate as compared to the other methods [14, 15].

In ELM, the hidden layer maps the input space to *L*-dimensional hidden layer feature space where, input weights ($a_{ij}$)) & hidden layer biases ($b_{0j}$) are chosen randomly. The structure of the ELM is shown in Figure 2.

The hidden layer output of '$j^{th}$' hidden layer for '$i^{th}$' pattern is calculated as in eq.(4).

$$h_j(X_i)_{\substack{j=1 \to L \\ i=1 \to S}} = \tanh\left( \sum_{k=1}^{n} (a_{ij} x_k) + b_{0j} \right) \qquad (4)$$

The total hidden layer output matrix can be written as in eq. (5)

$$[H]_{S \times L} = [h_1(X_i) \quad h_2(X_i) \quad \cdots \quad h_L(X_i)]_{S \times L} \qquad (5)$$

The goal of ELM is to achieve the smallest training error as well as the smallest norm of output weights (*β*) to achieve better generalization performance. This can be expressed as in eq. (6)

*Minimize*: $\| H_{trn}\beta - D_{trn} \|$ *and Minimize*: $\|\beta\|$ \qquad (6)

Where, '$H_{trn}$' is the hidden layer output matrix calculated from training input pattern '$X_{trn}$'. The smallest norm least square solution of the above linear system can be expressed as in eq. (7).

$$H_{trn}\beta = D_{trn} \Rightarrow [\beta]_{L \times m} = [H_{trn}^\dagger][D_{trn}] \qquad (7)$$

Where, '$H_{trn}^\dagger$' is the Moore-Penrose inverse of the non-singular matrix '$H_{trn}$' and thus eq. (7) can be modified as in eq. (8).

$$\beta = (H_{trn}^T H_{trn})^{-1} H_{trn}^T D_{train} \qquad (8)$$

Once the network is trained, the output weights are calculated and fixed which is later used to test the model performance. The predicted output ($O_{tst}$) by the network can be written as:

$$O_{tst} = H_{tst}\beta$$
$$\Rightarrow O_{tst} = H_{tst}H_{trn}^T (H_{trn}H_{trn}^T)^{-1} D_{trn} \qquad (9)$$

Where, '$H_{tst}$' is the hidden layer output matrix for testing patterns.

Even if ELM has various advantages still it suffers from ddisadvantages like presence of random input weight which requires has to be selected optimally. Thus an extended ELM (EELM) is applied for solar power prediction where there are no random input weights present. The EELM provides more stability & better prediction accuracy in solar power forecasting application.

*C. Extended Extreme Learning Machine (EELM)*

The EELM is the combination of FLNN and ELM which performs better as compared to the individual methods. The combination of both models provides more non-linearity and thus produces more accurate results.

In EELM the hidden layer outputs are calculated directly from the extended inputs after the functional expansion block (FEB). As the feature space is consist of non-linear mapping thus a linear model in the feature space behaves like a nonlinear one.

Then, EELM trains the network only using the non-linear inputs from which the '$\beta$' matrix is calculated. This '$\beta$' value is kept constant after the training section and used in the testing section to check the precision of the model by predicting the future data. The structure of the EELM is shown in Fig.3.

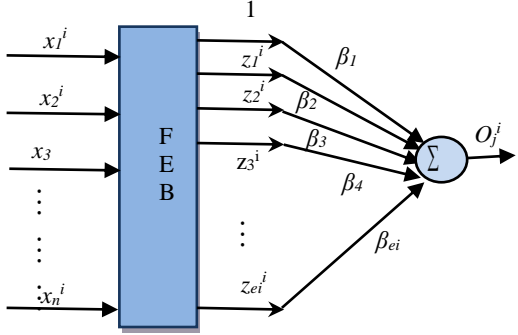

Figure 3. Structure of Extended ELM

The original input pattern ($X$) will be expanded by using the expansion rule explained in eq. (2). The total numbers of expanded input patterns are elements of the hidden layer matrix. Thus the hidden layer matrix for '$i^{th}$' input pattern can be written as in eq. (10).

$$h_j(X_i) = \begin{bmatrix} 1 & z_1^i & z_2^i & z_3^i & z_4^i & \cdots & z_{ei}^i \end{bmatrix}_{S \times (ei+1)} \quad (10)$$
$$i, j = 1 \rightarrow S$$

The total hidden layer matrix will be expressed as in eq. (11) where the number of hidden layers is the number of extended inputs.

$$[H]_{S \times (ei+1)} = [h_1(X_i) \quad h_2(X_i) \quad \cdots \quad h_S(X_i)]^T_{S \times (ei+1)} \quad (11)$$

Then, the output weight matrix and desired solar power calculation is same as eq. (8) and (9), respectively.

The above discussed models are used to predict the solar power in different weather condition and in different time horizon. The performance of the all the considered model is presented in the following section.

## III. RESULTS AND DISCUSSION

In this section, the superior performance of the proposed EELM is discussed against the basic ELM and FLNN techniques. Various case studies with figures with tables are presented for different weather situation. Further, the details of historical data are discussed with various performance metrics.

### A. Collection of Data

To predict the solar power from the past solar power data, the short term data (5 minutes and average 1 hour) are collected from the NREL [18]. The solar power plant is of 25 MW capacity and is located at Florida, USA, with latitude of 25.25 N and longitude of -80.85W. The data is collected from the 1st January 2006 to 31st December 2006. Then data acquisition for the considered models is obtained in seasonal manner (i.e. Summer season $\rightarrow$ month of March, April, May, June; 5.30 am−6.30 pm, Rainy season $\rightarrow$ month of July, August, September, October; 6.30 am−5.30 pm and Winter season $\rightarrow$ month of November, December, January, February; 8.00 am−4.00 pm).

### B. Performance Metrics

The performance of the proposed vs. existing prediction techniques is evaluated through different performance metrics like Root Mean Square Error (RMSE,eq.12), Symmetric Mean Absolute Percentage Error (SMAPE) and Mean Absolute Error (MAE, eq.13).

$$RMSE = \sqrt{\frac{\sum (D_{tst} - O_{tst})^2}{S_{tst}}} \quad (12)$$

$$MAE = \frac{\sum |D_{tst} - O_{tst}|}{S_{tst}} \quad (13)$$

As the solar power data contains zero value (null volatility), the SMAPE is considered to avoid infinity error in calculation. The relative error SMAPE [19] is expressed as in eq. (14).

$$SMAPE = \frac{\sum |D_{tst} - O_{tst}|}{\sum (D_{tst} + O_{tst})} \times 100 \quad (14)$$

Further for choosing the accurate model for solar power forecasting performance, the correlation coefficient ($CC^2$) is calculated. The $CC^2$ gives the amount of correlation of the measured values with the targeted values and defined as in eq. (15).

$$CC^2 = \frac{\left( S_{tst} \sum_{i=1}^{S_{tst}} (O(X_i) D_{ptst}(X_i)) \right)^2 - \sum_{i=1}^{S_{tst}} O(X_i) \sum_{i=1}^{S_{tst}} D_{tst}(X_i)}{\left( S_{tst} \sum_{i=1}^{S_{tst}} O(X_i)^2 - (\sum_{i=1}^{S_{tst}} O(X_i))^2 \right) \times \left( S_{tst} \sum_{i=1}^{S_{tst}} D_{tst}(X_i)^2 - (\sum_{i=1}^{S_{tst}} D_{tst}(X_i))^2 \right)} \quad (15)$$

The solar power is scaled and normalized within the range 0 and 1 as given by eq. (16).

$$\text{scaled } x(t) = \frac{x(t) - x_{\min}}{x_{\max} - x_{\min}} \quad (16)$$

Where, $x(t)$ represents the solar power at time instant $t$; $x_{\max}$ and $x_{\min}$ represent the maximum and minimum solar power.

### C. Results Analysis

Here two case studies are presented for 5 minute and 1 hour time horizon. In both the case, different weather data are considered to validate the superior performance of the proposed EELM model. Lastly, a comparative study is presented through bar graph representation in case-3.

*Case-1: Solar Power Prediction for 5 Minute Time Horizon*

The prediction results for 5 minute time horizon is as given in figure (4). The figures 4(b) and (c) show the results for rainy and winter season, respectively. As clearly seen from the figures the power prediction by the proposed EELM is the more accurate as compared with the other two considered methods.

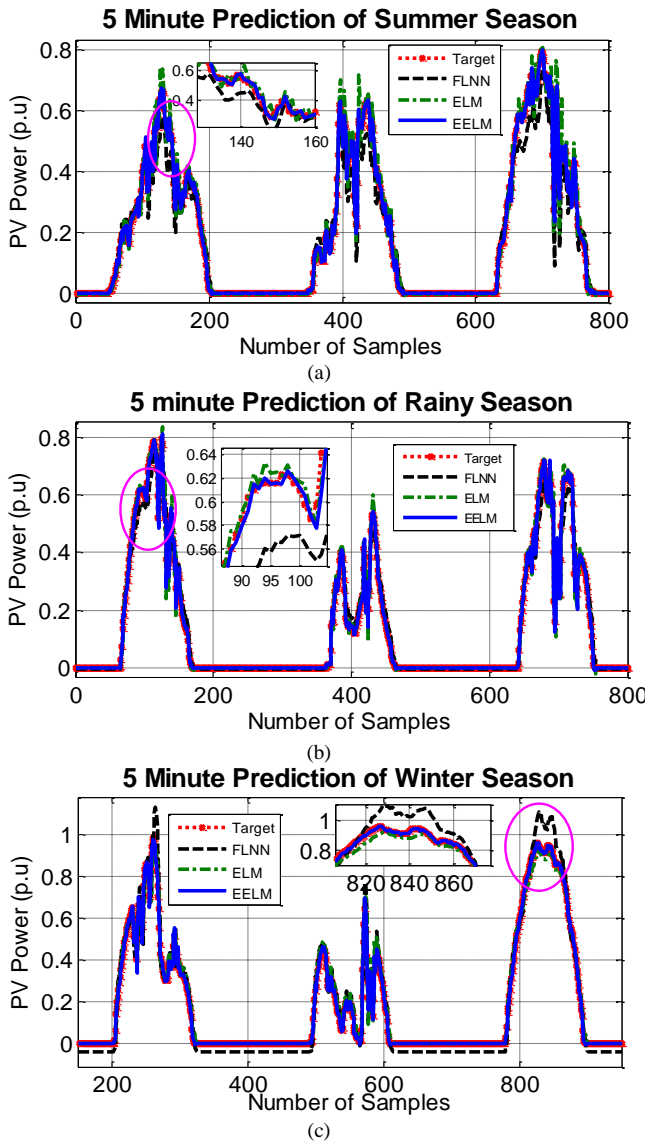

Figure 4. Performance analysis of prediction schemes for 5 minute time horizon: (a) Summer season, (b) Rainy season, (a) Winter season.

*Case-2: Solar Power Prediction for 1hour Time Horizon*

Similarly, the prediction results for 1 hour time horizon is as shown in figure (5). The figures 5(b) and (c) show the results for rainy and winter season, respectively. As clearly seen from the figures, the prediction accuracy of proposed EELM is the better as compared with the ELM and FLNN. As the number of samples are less for 1 hour time horizon the error are slightly higher as compared to the 5 minute prediction interval.

*Case-3: Performance validation through a Comparative Analysis*

The effectiveness of the proposed scheme for all seasonal data in terms of various performance indexes and training time (TT) is established in Tables 1 to 3 for different seasons. Table 1 shows the performance parameters for summer season. Table 2 and 3 show the performance parameters for rainy and winter season, respectively.

For summer season (Table 1), RMSE for proposed EELM is obtained as 0.0165 for 5 minutes interval and 0.0691 for 1 hour interval, where FLNN depicts 0.0368 and 0.1456 for the same. The training time for the proposed prediction scheme (5 minutes interval) is recorded as 0.09 sec which is less than FLNN (0.14 sec). The duration is more as compared with ELM scheme (0.05 sec.) because of its higher size of *H*-matrix. If both the execution time and accuracy factor will be considered then proposed EELM is better as compared to the basic ELM and FLNN techniques. The superior performance of the proposed EELM can also be observed for rainy and winter season from the table 2 and table 3, respectively. The performance indexes show that the proposed EELM performs better for different weather condition and different time horizon.

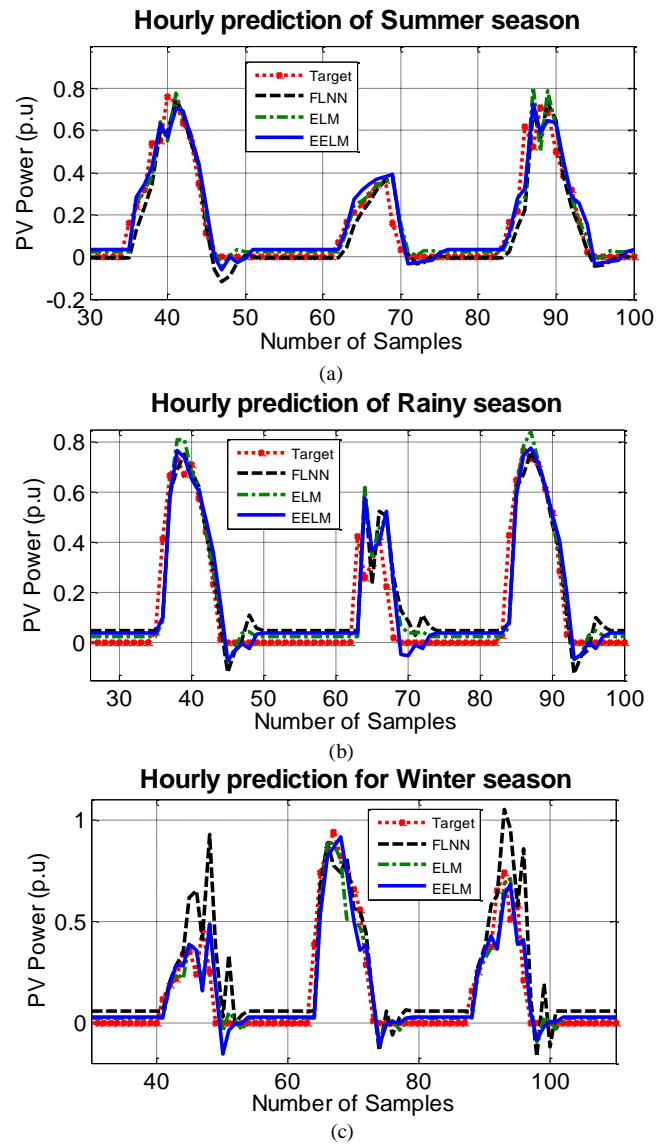

Figure 5. Performance analysis of prediction schemes for 1 hour time horizon: (a) Summer season, (b) Rainy season, (a) Winter season.

TABLE I. COMPARATIVE STUDY OF VARIOUS PREDICTION TECHNIQUES IN TERMS OF PERFORMANCE INDEX FOR SUMMER SEASON.

| SUMMER SEASON | | | | |
|---|---|---|---|---|
| Performance Index | Time Horizon | FLNN | ELM | EELM |
| RMSE (p.u) | 5 min. | 0.0368 | 0.0285 | **0.0165** |
| | 1 hour | 0.1456 | 0.0835 | **0.0691** |
| MAE (p.u) | 5 min. | 0.0213 | 0.0156 | **0.0071** |
| | 1 hour | 0.0985 | 0.0489 | **0.0262** |
| SMAPE (%) | 5 min. | 5.21 | 2.35 | **1.68** |
| | 1 hour | 19.32 | 14.38 | **13.52** |
| $CC^2$ | 5 min. | 0.9862 | 0.9909 | **0.9965** |
| | 1 hour | 0.9104 | 0.9239 | **0.9313** |
| TT (sec) | 5 min. | 0.14 | 0.05 | **0.09** |
| | 1 hour | 0.09 | 0.02 | **0.04** |

TABLE II. COMPARATIVE STUDY OF VARIOUS PREDICTION TECHNIQUES IN TERMS OF PERFORMANCE INDEX FOR RAINY SEASON.

| RAINY SEASON | | | | |
|---|---|---|---|---|
| Performance Index | Time Horizon | FLNN | ELM | EELM |
| RMSE (p.u) | 5 min. | 0.0300 | 0.0208 | **0.0164** |
| | 1 hour | 0.1045 | 0.0989 | **0.0792** |
| MAE (p.u) | 5 min. | 0.1456 | 0.0850 | **0.0081** |
| | 1 hour | 0.0743 | 0.0581 | **0.0435** |
| SMAPE (%) | 5 min. | 5.19 | 2.78 | **1.79** |
| | 1 hour | 19.55 | 17.99 | **15.50** |
| $CC^2$ | 5 min. | 0.9870 | 0.9911 | **0.9953** |
| | 1 hour | 0.8473 | 0.8668 | **0.8918** |
| TT (sec) | 5 min. | 0.18 | 0.06 | **0.07** |
| | 1 hour | 0.07 | 0.04 | **0.05** |

TABLE III. COMPARATIVE STUDY OF VARIOUS PREDICTION TECHNIQUES IN TERMS OF PERFORMANCE INDEX FOR WINTER SEASON.

| WINTER SEASON | | | | |
|---|---|---|---|---|
| Performance Index | Time Horizon | FLNN | ELM | EELM |
| RMSE (p.u) | 5 min. | 0.0417 | 0.0286 | **0.0158** |
| | 1 hour | 0.1456 | 0.1106 | **0.1048** |
| MAE (p.u) | 5 min. | 0.0932 | 0.0406 | **0.0063** |
| | 1 hour | 0.0985 | 0.0665 | **0.0357** |
| SMAPE (%) | 5 min. | 7.65 | 4.68 | **1.54** |
| | 1 hour | 16.67 | 14.38 | **12.21** |
| $CC^2$ | 5 min. | 0.9201 | 0.9571 | **0.9965** |
| | 1 hour | 0.8042 | 0.8932 | **0.9006** |
| TT (sec) | 5 min. | 0.17 | 0.045 | **0.086** |
| | 1 hour | 0.077 | 0.015 | **0.036** |

The supremacy of the proposed EELM can also be presented through a bar graph representation in figure (6). From figure (6), it is clearly seen that the prediction error in the proposed EELM is less as compared to the other two methods. Hence, the proposed EELM is efficient and effective to predict the future solar power which can be helpful in various power system aspects.

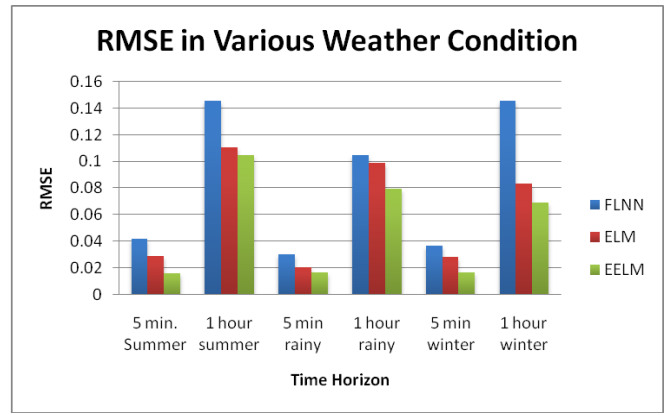

Figure 6. Bar graph representation of RMSE for various weather condition.

IV. CONCLUSION

A hybrid EELM technique is proposed in this paper to accurately predict the solar power. An online dataset of Florida for 5 minute and 1 hour time interval is considered for the proposed study. Various performance indexes like RMSE, SMAPE, MAE and $CC^2$ are calculated to prove the superiority of the EELM method. The performance of the EELM is compared with basic ELM and FLNN methods. The accuracy and effectiveness of the considered methods are validated through a comparative study for different weather condition. The main findings of this paper are (1) to develop a simple, fast yet efficient prediction model for solar power prediction, (2) the need of hidden layer selection and random input weights are avoided by using the proposed EELM approach. In future, this method can be implemented for the effective management of electricity and to increase the energy market efficiency.

REFERENCES

[1] Sobri, Sobrina, Sam Koohi-Kamali, and Nasrudin Abd Rahim. "Solar photovoltaic generation forecasting methods: A review." *Energy Conversion and Management* 156 (2018): 459-497.)

[2] Antonanzas, Javier, Natalia Osorio, Rodrigo Escobar, Ruben Urraca, Francisco J. Martinez-de-Pison, and Fernando Antonanzas-Torres. "Review of photovoltaic power forecasting." *Solar Energy* 136 (2016): 78-111.

[3] Lynch, Peter. "The origins of computer weather prediction and climate modeling." *Journal of Computational Physics* 227, no. 7 (2008): 3431-3444.

[4] Ren, Ye, P. N. Suganthan, and N. Srikanth. "Ensemble methods for wind and solar power forecasting—A state-of-the-art review." *Renewable and Sustainable Energy Reviews* 50 (2015): 82-91.

[5] Amini, M. Hadi, Amin Kargarian, and Orkun Karabasoglu. "ARIMA-based decoupled time series forecasting of electric vehicle charging demand for stochastic power system operation." *Electric Power Systems Research* 140 (2016): 378-390.

[6] Fentis, Ayoub, Lhoussine Bahatti, Mohammed Mestari, and Brahim Chouri. "Short-term solar power forecasting using Support Vector Regression and feed-forward NN." In *2017 15th IEEE International New Circuits and Systems Conference (NEWCAS)*, pp. 405-408. IEEE, 2017.

[7] Zeng, Jianwu, and Wei Qiao. "Short-term solar power prediction using a support vector machine." *Renewable Energy* 52 (2013): 118-127.

[8] Izgi, Ercan, Ahmet Öztopal, Bihter Yerli, Mustafa Kemal Kaymak, and Ahmet Duran Şahin. "Short–mid-term solar power prediction by using artificial neural networks." *Solar Energy* 86, no. 2 (2012): 725-733.

[9] Chen, S. X., H. B. Gooi, and M. Q. Wang. "Solar radiation forecast based on fuzzy logic and neural networks." *Renewable Energy* 60 (2013): 195-201.

[10] Sivaneasan, B., C. Y. Yu, and K. P. Goh. "Solar forecasting using ANN with fuzzy logic pre-processing." *Energy procedia* 143 (2017): 727-732.

[11] Eseye, Abinet Tesfaye, Jianhua Zhang, and Dehua Zheng. "Short-term photovoltaic solar power forecasting using a hybrid Wavelet-PSO-SVM model based on SCADA and Meteorological information." *Renewable Energy* 118 (2018): 357-367.

[12] Eseye, Abinet Tesfaye, Jianhua Zhang, and Dehua Zheng. "Short-term photovoltaic solar power forecasting using a hybrid Wavelet-PSO-SVM model based on SCADA and Meteorological information." *Renewable Energy* 118 (2018): 357-367.

[13] Wu, Yuan-Kang, Chao-Rong Chen, and Hasimah Abdul Rahman. "A novel hybrid model for short-term forecasting in PV power generation." *International Journal of Photoenergy* 2014 (2014).

[14] Bouzerdoum, Moufida, Adel Mellit, and A. Massi Pavan. "A hybrid model (SARIMA–SVM) for short-term power forecasting of a small-scale grid-connected photovoltaic plant." *Solar Energy* 98 (2013): 226-235.

[15] Satapathy, Prachitara, and Snehamoy Dhar. "A hybrid functional link extreme learning machine for Maximum Power Point Tracking of partially shaded Photovoltaic array." In *2015 IEEE Power, Communication and Information Technology Conference (PCITC)*, pp. 409-416. IEEE, 2015.

[16] Satapathy, Prachitara, Snehamoy Dhar, and P. K. Dash. "An evolutionary online sequential extreme learning machine for maximum power point tracking and control in multi-photovoltaic microgrid system." *Renewable Energy Focus* 21 (2017): 33-53.

[17] Satapathy, Prachitara, S. Dhar, and P. K. Dash. "A firefly optimized fast extreme learning machine based maximum power point tracking for stability analysis of microgrid with two stage photovoltaic generation system." *Journal of Renewable and Sustainable Energy* 8, no. 2 (2016): 025501.

[18] https://www.nrel.gov/grid/solar-power-data.html, Accessed on 15th January 2019.

[19] Shcherbakov, Maxim Vladimirovich, Adriaan Brebels, Nataliya Lvovna Shcherbakova, Anton Pavlovich Tyukov, Timur Alexandrovich Janovsky, and Valeriy Anatol'evich Kamaev. "A survey of forecast error measures." World Applied Sciences Journal 24 (2013): 171-176